\documentclass[graybox]{svmult}

\usepackage{type1cm}        
\usepackage{makeidx}         
\usepackage{graphicx}        
\usepackage{multicol}        
\usepackage[bottom]{footmisc}

\usepackage{newtxtext}       %
\usepackage[varvw]{newtxmath}       

\usepackage{xurl}
\usepackage[colorlinks,allcolors=blue]{hyperref}
\usepackage[authoryear]{natbib}

\usepackage{mathrsfs}%
\usepackage{booktabs}%
\usepackage{bm}%
\usepackage{enumitem}
\usepackage{threeparttable}

\makeindex             
							
\DeclareMathOperator{\Prob}{P}
\DeclareMathOperator{\Expec}{E}
\DeclareMathOperator{\Var}{Var}
\DeclareMathOperator{\Cov}{Cov}

\DeclareMathOperator{\II}{\mathbb{I}}

\newcommand{\toP}{\stackrel{\Prob}{\to}}
\newcommand{\toDSkorokhod}{\xrightarrow{\smash{\raisebox{-0.3ex}{$\scriptscriptstyle\mathcal{D}[0,1]$}}}}
\DeclareMathOperator{\diag}{diag}

\DeclareFontFamily{U}{mathx}{}
\DeclareFontShape{U}{mathx}{m}{n}{<-> mathx10}{}
\DeclareSymbolFont{mathx}{U}{mathx}{m}{n}
\DeclareMathAccent{\widehat}{0}{mathx}{"70}
\DeclareMathAccent{\widecheck}{0}{mathx}{"71}

\newcommand{\mat}[1]{\mathbf{#1}}
\renewcommand{\vec}[1]{\bm{#1}}
\renewcommand{\theta}{\vartheta}
\renewcommand{\hat}{\widehat}
\renewcommand{\check}{\widecheck}

\begin{document}

\title*{A Weighted Regression Approach to Break-Point Detection in Panel Data}
\author{Charl Pretorius and Heinrich Roodt}
\institute{Charl Pretorius \at Centre for Business Mathematics and Informatics, North-West University, Potchefstroom, South Africa \\\email{charl.pretorius@nwu.ac.za}
\and Heinrich Roodt \at Pure and Applied Analytics, North-West University, Potchefstroom, South Africa \\\email{heinrich.roodt.12@gmail.com}}
%
%
\maketitle

\abstract*{New procedures for detecting a change in the cross-sectional mean of panel data are proposed.
The procedures rely on estimating nuisance parameters using
certain cross-sectional means across panels using a weighted least squares regression.
In the case of weak cross-sectional dependence between
panels, we show how test statistics can be constructed to
have a limit null distribution not depending on any choice of
bandwidths typically needed to estimate the long-run variances of the panel errors.
The theoretical assertions are derived for general choices of the regression weights,
and it is shown that consistent test procedures can be obtained from the proposed process.
The theoretical results are extended to the case where strong cross-sectional
dependence exist between panels. The paper concludes with
a numerical study illustrating the behavior of several special cases
of the test procedure in finite samples.}

\abstract{New procedures for detecting a change in the cross-sectional mean of panel data are proposed.
The procedures rely on estimating nuisance parameters using
certain cross-sectional means across panels using a weighted least squares regression.
In the case of weak cross-sectional dependence between
panels, we show how test statistics can be constructed to
have a limit null distribution not depending on any choice of
bandwidths typically needed to estimate the long-run variances of the panel errors.
The theoretical assertions are derived for general choices of the regression weights,
and it is shown that consistent test procedures can be obtained from the proposed process.
The theoretical results are extended to the case where strong cross-sectional
dependence exist between panels. The paper concludes with
a numerical study illustrating the behavior of several special cases
of the test procedure in finite samples.}

\section{Introduction}\label{sec:intro}

We study the problem of detecting the presence of structural changes
in a panel data model in which there are $N$ panels (or variables), each containing
$T$ observations. Specifically, we consider the model
\begin{equation}\label{eq:mainmodel3}
		Y_{i,t} = \mu_i + \delta_i \II(t > t_0) + e_{i,t},
		\quad
		i=1,\ldots,N, \quad t=1,\ldots,T,
\end{equation}
in which the mean of panel $i$ changes from $\mu_i$ to $\mu_i+\delta_i$ at time
$t_0=\lfloor\theta T\rfloor$, for some unknown $\theta\in(0,1)$,
and $e_{i,t}$ is a zero-mean error.
Our main objective will be to study the properties of a test for a change in the mean
of at least one of the panels, i.e., a test for the hypothesis
\[
		H_0: \sum_{i=1}^N\delta_i^2 = 0
		\quad\text{versus}\quad
		H_A: \sum_{i=1}^N\delta_i^2 \ne 0.
\]

This paper focuses on test statistics based on functionals of
\begin{equation}\label{eq:VNT}\begin{split}
	V_{N,T}(s;\sigma^2)
	&= \frac{1}{\sqrt N}\sum_{i=1}^N \left(Z^2_{i,T}(s) - \sigma^2 m_T(s)\right),
	\quad
	m_T(s) = \frac{\lfloor Ts\rfloor}{T} \left(1 - \frac{\lfloor Ts\rfloor}{T}\right),
\end{split}\end{equation}
where $\sigma^2$ is chosen such that the process $V_{N,T}$ is properly centered, and
$Z_{i,T}$ denotes the CUSUM process calculated using observations from the $i$th panel, i.e.,
\begin{equation}\label{eq:cusumprocess}
		Z_{i,T}(s)
		= \frac{1}{\sqrt T}\sum_{k=1}^{\lfloor Ts \rfloor}\left( Y_{i,k} - \bar Y_{i,T} \right),
		\quad
		s\in(0,1),
\end{equation}
with $\bar Y_{i,T}=T^{-1}\sum_{k=1}^T Y_{i,k}$.
A possible test for the null hypothesis of no change rejects for large values of 
$\sup_{s\in(0,1)}\left|V_{N,T}(s;\sigma^2)\right|$ or
$\int_0^1 V_{N,T}^2(s;\sigma^2)ds$.

Our choice of $\sigma^2$ is based on the observation that, under $H_0$ and certain
regularity conditions,
\begin{equation}\label{eq:regr.idea}
\Expec\left(\frac{1}{N}\sum_{i=1}^N Z^2_{i,T}(s)\right)
  = \bar\sigma_N^2 m_T(s) + o(1), \quad
\Var\left(\frac{1}{N}\sum_{i=1}^N Z^2_{i,T}(s)\right)
  = \frac{2\bar\kappa_N^2}{N} m_T^2(s) + o(1),
\end{equation}
as $\min(N,T)\to\infty$, for some constants $\bar\sigma_N^2$ and
 $\bar\kappa_N^2$ possibly depending on $N$.
For instance, if the errors $e_{i,t}$ are independent across panels, 
form a linear process in each panel, and the long-run variances
$\sigma_i^2=\lim_{T\to\infty}\Expec(T^{-1/2}\sum_{t=1}^Te_{i,t})^2$,
$i=1,\ldots,N$,
exist, then \eqref{eq:regr.idea} holds with
$\bar\sigma_N^2=N^{-1}\sum_{i=1}^N\sigma_i^2$ and
$\bar\kappa_N^2=N^{-1}\sum_{i=1}^N\sigma_i^4$.

The relations in \eqref{eq:regr.idea} suggest that $\bar\sigma_{N}^2$ can be regarded
as the slope parameter in a linear regression of the observed cross-sectional means
of the squared CUSUM statistic on $m_T(\cdot)$. That is, we consider the regression
model
\begin{equation}\label{eq:regression}
    \frac{1}{N}\sum_{i=1}^N Z^2_{i,T}\left(\frac{k}{T}\right)
    = \bar\sigma_N^2 m_T\left(\frac{k}{T}\right) + m_T\left(\frac{k}{T}\right)\varepsilon_k,
    \quad
    k=1,\ldots,T,
\end{equation}
where the errors $\varepsilon_k$ are expected to have zero mean and constant variance.
The heteroscedastic model errors 
lead us to consider the weighted least squares estimator
\begin{equation}\label{eq:regression.estimator}
    \hat\sigma_{N,T}^2(\vec w_T)
		=\left(\vec m_T^\top\mat W_T\vec m_T^{}\right)^{-1}\vec m_T^\top\mat W_T\vec z_{N,T}^{},
\end{equation}
where $\mat W_T$ is a $(T-1)\times(T-1)$ diagonal matrix with
nonnegative entries $\vec w_t=\diag(\mat W_t)=(w_{1,T},\ldots,w_{T-1,T})\ne\vec 0$ on the diagonal, and
$\vec m_T$ and $\vec z_{N,T}$ are the $(T-1)$-vectors
\[\begin{split}
	\vec m_T&=\left[m_T\left(\frac{1}{T}\right),\ldots,m_T\left(\frac{T-1}{T}\right)\right]^\top,\\
	\vec z_{N,T}^{}&=\frac{1}{N}\sum_{i=1}^N \left[Z_{i,T}^2\left(\frac{1}{T}\right),\ldots,Z_{i,T}^2\left(\frac{T-1}{T}\right)\right]^\top.
\end{split}\]
Under $H_0$, if $(\vec m_T^\top\mat W_T^{}\vec m_T^{})^{-1}\vec m_T^\top\vec w_T^{}=o(T)$
and appropriate weak dependence conditions are imposed on the errors,
the quantity $\hat\sigma_{N,T}^2(\vec w_T)$
is an asymptotically unbiased estimator of $\bar\sigma_N^2$.
The unbiasedness implies that the mean of 
$\smash{V_{N,T}(\cdot;\hat\sigma_{N,T}^2(\vec w_T))}$
is approximately zero under $H_0$, so that the process
is properly centered.

A special case of the statistic in \eqref{eq:VNT} was studied recently by \citet{horvath2022}.
The authors consider $V_{N,T}(\cdot;\hat\sigma_{H,\tau}^2)$ with
$\hat\sigma_{H,\tau}^2:=N^{-1}\sum_{i=1}^N Z^2_{i,T}(\tau)/m_T(\tau)$,
for some fixed $\tau\in(0,1)$ chosen to control the size of the test.
\citet{horvath2022} show that $V_{N,T}(\cdot;\hat\sigma_{H,\tau}^2)$,
properly normalized, converges weakly
to a zero-mean Gaussian process. As the covariance structure of the limit
Gaussian process depends on an unknown nuisance parameter involving
the long-run variances $\sigma_i^2$, they propose a bootstrap approach
to estimate the null distribution in practical settings.
Sequential change-detection using 
related procedures is studied by \citet{huskova2025}.

Recently, several papers have appeared addressing the problem
of detecting changes in high-dimensional panel data. Our paper finds
its roots in the work of \citet{horvath2012} \citep[also see][]{chan2013},
which have since been extended in several directions.
\citet{antoch2019} consider
detection of changes in the intercept or slope of a panel regression model where
the regressors are also allowed to vary across panels.
\citet{huskova2024} consider the detection of mean-changes in the case where
the panels are allowed to depend on a common set of regressors and be cross-sectionally
dependent through a common factor model.
There we show that, under quite general conditions,
such as stationarity of the regressors, the test of \citet{horvath2012} can
be adapted in such a way that the limit null distribution of
the test does not depend on the regressors.
The detection of multiple change-points in panel data with cross-sectional dependence
is investigated by \citet{dueker2024}, who also consider the test of \citet{horvath2012},
among others. \citet{dueker2024} propose a wild block bootstrap method
to take cross-sectional and temporal dependence into account.

\citet{jo2021} propose a test for changes in the parameters in a dynamic
panel model containing observed and unobserved effects.
The case of estimating the break-point in this setting under long-range dependence
is treated by \citet{xi2025}.
\citet{zhao2024} introduce a new procedure based on signal statistics 
that can simultaneously identify multiple
change-points in sparse and dense high-dimensional data,
while being computationally efficient.

We briefly mention some other works related to change-point detection in a
multivariate setting.
\citet{huvskova2006ecf} develop and study the limit behavior of change-point
tests based on the empirical characteristic function (ecf); also see \citet{lee2022}
for a sequential procedure based on the ecf.
Also making use of
the ecf, \citet{hlavka2020} develop procedures
for paired and two-sample break-detection.
\citet{horvath2017} propose a CUSUM-type estimator of the time of change
in the mean of panel data in the presence of cross-sectional dependence
through a common factor model, and establish first- and second-order asymptotic
for inference.

An outline of the remainder of the paper is as follows.
In Section~\ref{sec:theory}, we study the asymptotic behavior of the process $V_{N,T}$
both under the null hypothesis and under the alternative.
In Section~\ref{sec:improvedestimator}, we propose an alternative estimator of
$\bar\sigma_N^2$ which is consistent also under the alternative.
Certain results are extended to the case of dependent panels
in Section~\ref{sec:dependent}.
Finally, the finite-sample behavior of the tests
is studied in Section~\ref{sec:simulations}.

\section{Theoretical results}\label{sec:theory}
The asymptotic behavior of the process $V_{N,T}$ under the null hypothesis
of no change as well as under the alternative will now be presented.
We focus on two specific cases: ordinary least squares and heteroscedasticity
weighted least squares.
Only the main results are presented and all proofs are deferred to the Appendix.

We assume that the underlying errors
in all panels are generated by independent strictly stationary strong mixing processes.
This allows for a wide range of data generating process, such
as certain ARMA and GARCH models which have become popular in the
applied time series literature.

Define the mixing rate $\alpha(\cdot)$ of a sequence $\{y_t,t\in\mathbb{Z}\}$ by
\[\label{eq:mixing.coeff}
	\alpha(k)
	= \sup_{n\in\mathbb{Z}} \sup_{A\in\mathcal{F}_{-\infty}^n, B\in\mathcal{F}_{n+k}^\infty} |\Prob(A \cap B) - \Prob (A) \Prob (B)|,
\]
where $\mathcal{F}_a^b$ denotes the $\sigma$-field generated by $\{y_t: a\le t\le b\}$.
A sequence with mixing rate $\alpha(\cdot)$ is said to be \emph{$\alpha$-mixing} (or \emph{strong mixing})
if $\alpha(k)\to 0$ as $k\to\infty$.

\begin{enumerate}[label=(A\arabic*),leftmargin=*]
	\item \label{assumption:moments}
				The $N$ error sequences $\{e_{i,t}, -\infty < t < \infty\}$, $i\in\{1,\ldots,N\}$,
				are strictly stationary and mutually independent.
				Furthermore, there exist finite constants $c_1,c_2,\Delta>0$ and $\nu>4$ such that
				\[ 
					\Expec e_{i,0} = 0,
					\quad
					\Expec |e_{i,0}|^{\nu+\Delta}<\infty,
					\quad\text{and}\quad
					c_1 \le \lim_{T\to\infty} \frac{1}{T} \Expec \left( \sum_{t=1}^T e_{i,t} \right)^2 \le c_2.
				\] 
	\item \label{assumption:mixing}
				The sequence $\{\vec e_t, -\infty<t<\infty\}$, where $\vec e_t=(e_{1,t},\ldots,e_{N,t})$,
				is assumed to be strong mixing with
				mixing coefficient $\alpha(\cdot)$ satisfying $M_{\rho,\Delta}(\alpha)<\infty$ for some even integer $\rho\ge\nu$, where
				$
					M_{\rho,\Delta} (\alpha) = \sum_{k=1}^\infty (k+1)^{\rho-2} \alpha(k)^{\Delta/(\rho+\Delta)}.
				$
\end{enumerate}

\subsection{Behavior under the null hypothesis}
To facilitate exposition, we introduce some additional notation.
Let $\mat C_T$ be the $(T-1)\times(T-1)$ matrix with entry
$(k,\ell)$ set equal to $g^2(k/T,\ell/T)$, where $g(s,t)=(s\wedge t)(1-s\vee t)$.
Also define the symmetric function
\begin{equation}\label{eq:covkernel}
	\gamma(s,t\,|\,D,h)
	= 2 \left\{g^2(s,t)
		- h(s)m(t)
		- h(t)m(s)
		+ Dm(s)m(t)\right\},
	\quad
	s,t\in(0,1),
\end{equation}
where $m(s)=s(1-s)$,
\begin{equation}\begin{split}\label{eq:Dh}
	D &= \lim_{T\to\infty}D_T(\vec w_T) := \lim_{T\to\infty}\left(\vec m_T^\top\mat W_T^{}\vec m_T^{}\right)^{-2}\vec m_T^\top\mat W_T^{}\mat C_T^{}\mat W_T^{}\vec m_T^{}, \\
	h(s) &= \lim_{T\to\infty}h_T(s;\vec w_T) := \lim_{T\to\infty}\left(\vec m_T^\top\mat W_T^{}\vec m_T^{}\right)^{-1}\sum_{k=1}^{T-1}g^2\left(\frac{\lfloor Ts\rfloor}{T},\frac{k}{T}\right)m_T\left(\frac{k}{T}\right)w_{k,T},
\end{split}\end{equation}
and it is supposed that these limits exist.
%

The limit null distribution of $V_{N,T}$ is stated in the following theorem,
where $\toDSkorokhod$ denotes weak convergence in the Skorokhod space.
Recall that $\bar\kappa_N^2=N^{-1}\sum_{i=1}^N\sigma_i^4$, with
$\sigma_i^2=\lim_{T\to\infty}\Expec(T^{-1/2}\sum_{t=1}^Te_{i,t})^2$,
$i=1,\ldots,N$, and let $\bar\kappa^2=\lim_{N\to\infty}{\bar\kappa}_N^2$.
\begin{theorem}\label{thm1}
Suppose Assumptions \ref{assumption:moments} and \ref{assumption:mixing} hold, and that
\begin{equation}\label{thm1.conditions}
	\frac{\sqrt N}{T}\to0
	\qquad\text{and}\qquad
	\eta_T := \frac{\vec m_T^\top\vec w_T^{}}{\vec m_T^\top\mat W_T^{}\vec m_T^{}} = o\left(\frac{T}{\sqrt N}\right)
\end{equation}
as $\min(N,T)\to\infty$.
Then, under $H_0$,
\[V_{N,T}(\cdot;\hat\sigma_{N,T}^2(\vec w_T))\toDSkorokhod \bar\kappa G,\]
where $G$ is a zero-mean Gaussian process with covariance kernel
$\gamma(s,t\,|\,D,h)$.
\end{theorem}

Notice that the exact limit behavior of the process
$V_{N,T}$ depends on the unknown quantity $\bar\kappa$
which needs to be estimated. Estimation of $\bar\kappa$ will
be discussed in Section~\ref{sec:kappaest}.

We consider three special cases resulting from specific choices of $\vec w_T$.
Firstly, suppose $\vec w_T$ is chosen such that the regression in \eqref{eq:regression.estimator}
is ordinary least squares regression.
That is, we consider the process $\hat V_{N,T}^{\rm ols}(\cdot):=V_{N,T}(\cdot;\hat\sigma_{N,T}^2(\vec w_T^{\rm ols}))$,
where $\vec w_T^{\rm ols}$ is a $(T-1)$-vector consisting of ones. In this case, the following holds.
\begin{corollary}\label{thm.ols}
Suppose Assumptions \ref{assumption:moments} and \ref{assumption:mixing} hold
and that $\sqrt N T^{-1}\to 0$.
Then, under $H_0$,
$\hat V_{N,T}^{\rm{ols}}\toDSkorokhod \bar\kappa G_{\rm{ols}}$ as $\min(N,T)\to\infty$,
where $G_{\rm{ols}}$ is a zero-mean Gaussian process with covariance kernel
$\gamma(s,t\,|\,D_{\rm{ols}},h_{\rm{ols}})$ and
\[
	D_{\rm{ols}}=\frac{13}{28},
	\quad
	h_{\rm{ols}}(s)=\frac{3}{2}m^2(s)\left(1+2m(s)\right).
\]
\end{corollary}

For the second case, namely weighted least squares regression, where the
weights are chosen according to the form of heteroscedasticity
present in \eqref{eq:regression}, the following result can be obtained.
Here, $\hat V_{N,T}^{\rm{wls}}(\cdot):=V_{N,T}(\cdot;\hat\sigma_{N,T}^2(\vec w_T^{\rm wls}))$,
where $\vec w_T^{\rm wls}$ is a $(T-1)$-vector with $k$th entry equal to $m^{-2}(k/T)$.
Notice that, because of the heavy weight placed near the boundaries of the interval $(0,1)$,
a stronger condition on the relation between the number of
panels and number of observations is needed.
\begin{corollary}\label{thm.wls}
Suppose Assumptions \ref{assumption:moments} and \ref{assumption:mixing} hold
and that $\sqrt N T^{-1}\log T\to 0$.
Then, under $H_0$,
$\hat V_{N,T}^{\rm{wls}}\toDSkorokhod \bar\kappa G_{\rm{wls}}$ as $\min(N,T)\to\infty$,
where $G_{\rm{wls}}$ is a zero-mean Gaussian process with covariance kernel
$\gamma(s,t\,|\,D_{\rm{wls}},h_{\rm{wls}})$ and
\[
	D_{\rm{wls}}=\frac{1}{3}\pi^2-3,
	\quad
	h_{\rm{wls}}(s)=-\left[s^2\log s + (1-s)^2\log(1-s) + m(s)\right].
\]
\end{corollary}
Under the conditions of Corollary~\ref{thm.wls}, one has
$\eta_T=2\log T+O(1)$. Therefore, the criteria $\sqrt N T^{-1}\log T\to 0$ is
needed to satisfy \eqref{thm1.conditions}.

Lastly, if $\vec w_T$ is chosen such that $w_{k,T}=\II(k=\lfloor\tau T\rfloor)$ for some $\tau\in(0,1)$,
one obtains the test of \citet{horvath2022}. Denoting this vector by $\smash{\vec w_T^{(\tau)}}$, one obtains
the estimator
$\smash{\hat\sigma_{H,\tau}^2:=\hat\sigma_{N,T}^2(\vec w_T^{(\tau)})=N^{-1}\sum_{i=1}^N Z^2_{i,T}(\tau)/m_T(\tau)}$, $\tau\in(0,1)$,
defined in the introduction. 
If $H_0$ holds and $\sqrt N T^{-1}\to 0$ as $\min(N,T)\to\infty$, then $V_{N,T}(\cdot;\hat\sigma_{H,\tau}^2)\toDSkorokhod\bar\kappa G_\tau$,
where $G_\tau$ is a zero-mean Gaussian process with covariance kernel $\gamma(s,t\,|\,1,h_\tau)$ and
\[
	h_\tau(s)=\frac{g^2(s,\tau)}{m(\tau)}=\frac{\left(s\wedge\tau\right)^2\left(1-s\vee\tau\right)^2}{\tau(1-\tau)}.
\]
In this case, the distribution of $G_\tau$ stated here coincides with that given in
Theorem~2.1 of \citet{horvath2022}.

\begin{remark}
The condition $\sqrt{N}T^{-1}\to 0$ allows the number of panels $N$ to be larger (asymptotically)
than the number of observations $T$. This agrees with the condition specified in
\citet{horvath2012} and is necessary for Corollary~\ref{thm.ols} to hold.
More recently, \citet{horvath2022} have shown that, if the errors in each panel
are serially uncorrelated,
the process $V_{N,T}(\cdot;\hat\sigma_{H,\tau}^2)$
converges weakly to a Gaussian process, even if the condition $\sqrt{N}T^{-1}\to 0$ is dropped.
Careful inspection of the proofs show that this condition may also be dropped from our
Corollary~\ref{thm.ols} if the errors are serially uncorrelated.
In fact, in Theorem~3.6 of \citet{horvath2022} one sees that this condition has to be
imposed on the relation between $N$ and $T$ in the case of serially correlated errors.
\end{remark}

\subsection{Behavior under the alternative}
In the case that a change-point is present in the observed data, and the change is large enough,
a test based on $\sup_{s\in(0,1)}|V_{N,T}(s;\hat\sigma_{N,T}^2(\vec w_T))|$ is able to detect the
change with high probability. This claim is made precise in Theorem~\ref{thm:VNT.consistency}
below.

\begin{theorem}\label{thm:VNT.consistency}
	Suppose Assumptions~\ref{assumption:moments} and \ref{assumption:mixing} hold.
	If $\sqrt NT^{-1}\to0$, $\eta_T=o(TN^{-1/2})$, and
	\[
		\frac{T}{\sqrt N}\sum_{i=1}^N\delta_i^2\to\infty
	\]
	as $\min(N,T)\to\infty$, then
	\[
		\sup_{s\in(0,1)}\left|V_{N,T}\left(s;\hat\sigma_{N,T}^2(\vec w_T)\right)\right|\toP\infty.
	\]
\end{theorem}

In Section~\ref{sec:improvedestimator} we study the behavior under the alternative
more closely.

\subsection{Estimation of the remaining nuisance parameter}\label{sec:kappaest}
As estimator of the nuisance parameter $\bar\kappa$ appearing in the limit
null distribution, we follow the same regression-based idea. Under the stated
assumptions (with a stronger moment condition $\nu\ge8$), it can be shown that
\[\begin{split}
	\Expec\left(\frac{1}{N}\sum_{i=1}^N Z^4_{i,T}(s)\right)
    = 3\bar\kappa_N^2 m_T^2(s) + o(1).
\end{split}\]
One can therefore estimate $\bar\kappa_N^2$ using the least-squares estimator
\begin{equation}\label{eq:regression.estimator.kappa}
    \hat\kappa_{N,T}^2(\vec w_T)=\frac{1}{3}\left(\ddot{\vec m}_T'\mat W_T^{}\ddot{\vec m}_T^{}\right)^{-1}\ddot{\vec m}_T'\mat W_T^{}\ddot{\vec z}_{N,T}^{},
\end{equation}
where $\mat W_T=\diag(\vec w_T)$ is the diagonal matrix defined earlier,
$\ddot{\vec m}_T$ is a vector with $k$th element equal to $m_T^2(k/T)$, $k=1,\ldots,T-1$,
 and $\ddot{\vec z}_T$ a vector with $k$th element equal to $N^{-1}\sum_{i=1}^N Z_{i,T}^4(k/T)$.
\begin{theorem}\label{theorem:consistency_kappa1}
	Suppose that Assumptions~\ref{assumption:moments} and \ref{assumption:mixing} hold
	with $\nu\ge 8$.
	If
	\[
		\frac{\sqrt{N}}{T}\to 0
		\qquad\text{and}\qquad
		\ddot{\eta}_T:=\frac{\ddot{\vec m}_T^\top\vec w_T^{}}{\ddot{\vec m}_T^\top\mat W_T^{}\ddot{\vec m}_T^{}} = o\left(\frac{T}{\sqrt N}\right)
	\]
	then, under $H_0$,
	\[\begin{split}
		\Expec\left(\hat\kappa_{N,T}^2(\vec w_T) - \bar\kappa_N^2\right) = o(1)
		\qquad\text{and}\qquad
		\Expec\left(\hat\kappa_{N,T}^2(\vec w_T) - \bar\kappa_N^2\right)^2 = o(1),
	\end{split}\]
	as $\min(N,T)\to\infty$.
\end{theorem}

Theorem~\ref{theorem:consistency_kappa1} implies that, under the stated conditions,
$\smash{\hat\kappa_{N,T}^2(\vec w_T)}$ is a consistent estimator of $\bar\kappa^2$.
Therefore, in conjunction with Theorem~\ref{thm1}, whenever the null hypothesis is true,
\[
    \frac{1}{\hat\kappa_{N,T}(\vec w_T)} \hat{V}_{N,T} \toDSkorokhod G.
\]
The quantity on the right-hand side is pivotal as it depends on no unknown
parameters, which renders the test criterion on the left-hand side suitable for practical application.

\subsection{Numerical study}\label{numstudy.asymp}
We now illustrate the behavior of tests based on the criterion $\sup_{s\in(0,1)}|V_{N,T}(s;\vec w_T)|$ when
using the asymptotic critical values (at the 5\% significance level) implied by
Theorem~\ref{thm1}. The critical values are approximated
by means of Monte Carlo simulation, using 10\,000 independent sample paths
of the Gaussian process $G$ evaluated at 1\,000 points in $(0,1)$.
Four different choices of the weight vector $\vec w_T$ in \eqref{eq:regression.estimator}
and \eqref{eq:regression.estimator.kappa} are considered,
namely, $\vec w_T^{\rm{ols}}$ used in Corollary~\ref{thm.ols},
 $\vec w_T^{\rm{wls}}$ used Corollary~\ref{thm.wls}, and $\smash{\vec w_T^{(\tau)}}$ for $\tau=0.1,0.5$.
The tests based on these weights are referred to by $\hat V^{\rm{ols}}$, $\hat V^{\rm{wls}}$,
$\hat V^{0.1}$, and $\hat V^{0.5}$, respectively.

Table~\ref{tab:asymp} shows the rejection percentages when panel data are simulated
according to \eqref{eq:mainmodel3} with the errors $e_{i,t}$ chosen as one of the following,
with $\varepsilon_{i,t}$ denoting i.i.d.\ N$(0,1)$ random variables:
\begin{enumerate}[label=(M\arabic*),leftmargin=*]
	\item \label{M1}
				the AR$(1)$ process $e_{i,t}=\rho e_{i,t-1} + \varepsilon_{i,t}$ with $|\rho|<1$,
				referred to in the tables as AR$(\rho)$;
	\item \label{M2}
				the ARMA$(2,1)$ process $e_{i,t}=0.2e_{i,t-1}-0.3e_{i,t-2}+\varepsilon_{i,t}+0.2\varepsilon_{i,t-1}$,
				referred to in the tables as ARMA.
\end{enumerate}
The left-hand panel of Table~\ref{tab:asymp} shows the rejection percentages
when data is generated under the null hypothesis. The right-hand panel
corresponds to the case where there is a change in the mean at time
$t_0=\lfloor\frac{1}{2}T\rfloor$
in 50\% of panels. For each panel with a change, the size $\delta_i$ of the change is drawn randomly from a uniform distribution on $[-0.4,0.4]$.

In most cases, the tests $\hat V^{\rm{ols}}$ and $\hat V^{\rm{wls}}$ have
reasonable empirical size close to the 5\% nominal level. The tests based on
$\hat V^{0.1}$ and $\hat V^{0.5}$ tend to be liberal for smaller samples,
which is likely due to the variability associated with estimating the mean long-run
variance at a single time-point $\lfloor\tau T\rfloor$. Nevertheless, as $N$ and $T$
increase, the empirical size of all tests seem to tend to the nominal level.

Comparing the power obtained by the tests $\hat V^{\rm{ols}}$ and $\hat V^{\rm{wls}}$,
we see that there is a significant increase in power when the weight vector $\vec w_T$
is weighted according to the heteroscedasticity mentioned earlier.
The test $\smash{\hat V^{0.1}}$ seems to have the highest power in most cases considered,
whereas the test $\smash{\hat V^{0.5}}$ has the lowest. This corresponds with
the recommendation of \citet{horvath2022} to not choose $\tau$ too close to the
true change-point.

\begin{table}
	\centering
	\caption{Rejection percentages at the $5\%$ significance level using asymptotic critical
					values.
					The left-hand panel corresponds to the case of no change, whereas the right-hand
					panel corresponds to the case where random changes occur in 50\% of the panels.}
	\label{tab:asymp}
	\begin{tabular}{lrrrrrrrrrr}
	\toprule
	&&& \multicolumn{4}{c}{$\delta_i=0$} & \multicolumn{4}{c}{$\delta_i\sim\operatorname{U}[-0.4,0.4]$} \\
	\cmidrule(lr){4-7}\cmidrule(lr){8-11}
	Model & $N$ & $T$ & $\hat V^{\rm{ols}}$ & $\hat V^{\rm{wls}}$ & $\hat V^{0.1}$ & $\hat V^{0.5}$
										& $\hat V^{\rm{ols}}$ & $\hat V^{\rm{wls}}$ & $\hat V^{0.1}$ & $\hat V^{0.5}$ \\
	\midrule
	AR(0) & 50 & 50 & 4.8 &  3.2 & 11.6 &  9.4 &  3.7 &  10.6 &  24.7 &  1.8 \\
	 &  & 100 & 5.4 &  3.4 & 10.1 &  8.3 &  6.4 &  32.9 &  47.5 &  0.9 \\
	 &  & 200 & 5.4 &  4.0 & 11.8 & 11.9 & 22.7 &  80.2 &  83.5 &  2.2 \\
	\cmidrule{2-11}
	 & 100 & 50 & 3.4 &  2.9 &  7.2 &  5.9 &  4.3 &  19.3 &  31.7 &  2.1 \\
	 &  & 100 & 3.6 &  3.2 &  6.7 &  5.7 & 22.5 &  66.4 &  68.8 &  5.5 \\
	 &  & 200 & 5.0 &  3.6 &  8.3 &  7.7 & 73.5 &  99.2 &  97.2 & 30.0 \\
	\cmidrule{2-11}
	 & 200 & 50 & 2.3 &  1.6 &  6.6 &  5.2 & 11.4 &  36.8 &  44.3 &  7.3 \\
	 &  & 100 & 4.2 &  2.3 &  5.5 &  4.6 & 65.8 &  94.7 &  91.7 & 41.4 \\
	 &  & 200 & 4.1 &  2.9 &  6.3 &  6.0 & 99.3 & 100.0 & 100.0 & 95.4 \\
	\midrule
	AR(0.3) & 50 & 50 & 2.6 &  5.5 & 16.8 &  6.8 &  1.8 &   8.8 &  23.8 &  2.0 \\
	 &  & 100 & 3.7 &  4.4 & 11.4 &  8.3 &  2.2 &  14.8 &  29.2 &  0.9 \\
	 &  & 200 & 4.0 &  4.7 & 10.9 &  8.7 &  6.1 &  37.9 &  50.1 &  0.6 \\
	\cmidrule{2-11}
	 & 100 & 50 & 2.0 &  5.2 & 14.3 &  4.1 &  2.1 &  15.4 &  29.0 &  1.1 \\
	 &  & 100 & 3.5 &  4.5 &  8.7 &  6.1 &  4.5 &  30.8 &  38.3 &  1.8 \\
	 &  & 200 & 4.3 &  5.1 &  9.4 &  6.8 & 20.8 &  72.4 &  68.3 &  7.4 \\
	\cmidrule{2-11}
	 & 200 & 50 & 1.2 & 10.7 & 17.5 &  2.5 &  2.5 &  34.2 &  44.6 &  2.0 \\
	 &  & 100 & 2.2 &  7.1 &  9.5 &  3.3 & 15.3 &  61.1 &  57.6 & 10.6 \\
	 &  & 200 & 3.5 &  4.4 &  6.4 &  5.5 & 61.0 &  97.1 &  93.0 & 41.2 \\
	\midrule
	ARMA & 50 & 50 & 6.2 &  2.5 &  6.1 & 10.0 &  3.6 &   5.5 &  13.2 &  2.8 \\
	 &  & 100 & 6.6 &  2.0 &  7.4 & 10.8 &  5.3 &  19.1 &  34.2 &  1.1 \\
	 &  & 200 & 7.7 &  3.9 &  8.8 & 14.3 & 17.4 &  63.7 &  72.4 &  1.9 \\
	\cmidrule{2-11}
	 & 100 & 50 & 3.6 &  2.1 &  4.1 &  7.7 &  2.5 &   5.3 &  13.4 &  1.9 \\
	 &  & 100 & 3.9 &  2.0 &  5.1 &  8.3 & 13.4 &  39.9 &  45.1 &  6.0 \\
	 &  & 200 & 5.9 &  3.8 &  4.4 &  9.2 & 60.3 &  94.3 &  90.4 & 22.7 \\
	\cmidrule{2-11}
	 & 200 & 50 & 3.0 &  1.9 &  2.7 &  5.8 &  5.8 &   9.1 &  16.7 &  4.2 \\
	 &  & 100 & 3.2 &  2.0 &  3.9 &  6.2 & 40.0 &  72.5 &  68.7 & 27.3 \\
	 &  & 200 & 5.6 &  4.0 &  7.5 &  6.6 & 97.4 & 100.0 &  99.2 & 85.9 \\
	\bottomrule
	\end{tabular}
\end{table}

\section{Alternative estimator of the mean long-run variance}\label{sec:improvedestimator}
It can be shown that, under the stated assumptions, the estimator $\hat\sigma_{N,T}^2$
defined in \eqref{eq:regression.estimator} typically is a consistent and asymptotically
unbiased estimator of $\bar\sigma_N^2$ whenever the null hypothesis of no change holds.
Despite this, one can show that its value tends to be inflated in the presence of a structural break.
The effect is a potential reduction in the power of the test.
We now illustrate the behavior of this estimator under the alternative
and study an improved estimator.

Consider again the panel model
defined in \eqref{eq:mainmodel3} and suppose that $\delta_i\ne 0$ for at least some $i$.
Under proper conditions,
\[\begin{split}
    \sqrt{N}\Expec\hat\sigma_{N,T}^2(\vec w_T)
    ={}&{} \sqrt{N}\bar\sigma_N^2+h_T(\vartheta;\vec w_T)\frac{T}{\sqrt N}\sum_{i=1}^N\delta_i^2 + o\left(\frac{T}{\sqrt N}\sum_{i=1}^N\delta_i^2\right),
\end{split}\]
with $h_T$ as defined in \eqref{eq:Dh}.
$\Expec\hat\sigma_{N,T}^2>\bar\sigma_N^2$ under $H_A$ for large enough $N$ and $T$
if one assumes that $TN^{-1/2}\sum_{i=1}^N\delta_i^2\to\infty$.
This means that, even if the process $\hat V_{N,T}(u)$ is evaluated at the true change-point $u=\vartheta$,
the second term in $\hat V_{N,T}(\vartheta)$ is overestimated, leading to a potential decrease
in the value of $\hat V_{N,T}(u)$ resulting in a loss of power.
It can be shown that, under $H_A$, one has $\Prob(\hat\sigma_{N,T}^2>\bar\sigma_N^2)\to1$
as $\min(N,T)\to\infty$; see~\eqref{consistency.term2}.

An alternative estimator of $\bar\sigma_N^2$ will now be introduced.
Define the centered panels 
\[
	\check Y_{i,t}(u) = Y_{i,t} - \hat\delta_{i,T}\left(u\right)\II(t>\lfloor Tu\rfloor),
	\quad
	i=1,\ldots,N,
	\quad
	t=1,\ldots,T,
	\quad
	u\in(0,1),
\]
where $\hat\delta_{i,T}\left(u\right)$ is an estimator of the size of a possible change
at $\lfloor Tu\rfloor$ in panel~$i$. Specifically,
\begin{equation}\label{eq:delta.hat}
	\hat\delta_{i,T}\left(u\right) = \frac{1}{T-\left\lfloor Tu \right\rfloor}\sum_{t=\left\lfloor Tu \right\rfloor+1}^T Y_{i,t} - \frac{1}{\left\lfloor Tu \right\rfloor}\sum_{t=1}^{\left\lfloor Tu \right\rfloor} Y_{i,t}.
\end{equation}
Define the CUSUM process $\check Z_{i,T}$ based on the centered panel data as
\[
	\check Z_{i,T}\left(s,u\right) = \frac{1}{\sqrt T}\sum_{t=1}^{\lfloor Ts\rfloor}\left(\check Y_{i,t}(u) - \frac{1}{T}\sum_{\ell=1}^T \check Y_{i,\ell}(u)\right),
	\qquad
	u,s\in(0,1).
\]
Now, suppose that $H_A$ is true and there is a change at $\lfloor\theta T\rfloor$. Under the stated assumptions,
\begin{equation}\begin{split}\label{eq:expec.Zhatsq}
	\Expec\left[\check Z_{i,T}^2(s,u)\right]
	={}&{} \sigma_i^2\left[m_T(s) - \frac{g_T^2(s,u)}{m_T(u)}\right] \\
			{}&{} + \delta_i^2T\left[g_T(s,\vartheta)-\frac{g_T(s,u)g_T(\theta,u)}{m_T(u)}\right]^2 + O(T^{-1}),
\end{split}\end{equation}
where
$g_T(s,u) = g({\lfloor Ts\rfloor}/{T},{\lfloor Tu\rfloor}/{T})$.
It is important to realize that \eqref{eq:expec.Zhatsq} holds \emph{even if the null hypothesis is violated}.
Moreover, if $\Expec[\check Z_{i,T}^2(s,u)]$ is evaluated with $uT\in[\lfloor\theta T\rfloor,\lfloor\theta T\rfloor+1)$,
i.e.\ with $u$ sufficiently close to $\theta$, the second term in the right-hand side of \eqref{eq:expec.Zhatsq} disappears.

Using the same regression idea, we base a test for $H_0$ on the process
$\check{V}_{N,T}(u):=V_{N,T}(u;\check\sigma_{N,T}^2(u))$, with
$\check\sigma_{N,T}^2(u) =(\check{\vec m}_T^\top\check{\vec m}_T^{})^{-1}\check{\vec m}_T^\top\check{\vec z}_{N,T}^{}(u)$,
where
\[\begin{split}
	\check {\vec z}_{N,T}^{}(u)&=\frac{1}{N}\sum_{i=1}^N \left[\check{Z}_{i,T}^2\left(\frac{1}{T},u\right),\ldots,\check{Z}_{i,T}^2\left(\frac{T-1}{T},u\right)\right]^\top
\end{split}\]
and, motivated by the first term in \eqref{eq:expec.Zhatsq},
$\check{\vec m}_T$ is a $(T-1)$-vector with $k$th entry set equal to
\[
	m_T(k/T) - \frac{g_T^2(k/T,u)}{m_T(u)},
	\quad
	k=1,\ldots,T-1.
\]

It can be shown that, whenever the null hypothesis of no change holds,
$\check\sigma_{N,T}^2(u)$ is an asymptotically unbiased and consistent
estimator of $\bar\sigma_N^2$, regardless the value of $u\in(0,1)$.
On the other hand, if a change is present, then the estimator is generally not consistent and unbiased.
However, if the estimator $\check\sigma_{N,T}^2(\cdot)$ is evaluated in
a small enough neighborhood of the true change-point $\theta$,
the estimator is asymptotically unbiased and consistent also under the alternative.
Below we formulate these results.
\begin{theorem}\label{thm:consistency.check.sigma2}
	Suppose Assumptions \ref{assumption:moments} and \ref{assumption:mixing} hold. If $H_0$ is true, then
	\[
		\sup_{u\in(0,1)}\left|\Expec\left(\check\sigma_{N,T}^2\left(u\right) - \bar\sigma_N^2\right)\right|\to0
	\qquad\text{and}\qquad
		\sup_{u\in(0,1)}\left|\check\sigma_{N,T}^2\left(u\right) - \bar\sigma_N^2\right| \toP 0.
	\]
	If $H_A$ is true, then
	\[
		\sup_{u\in B_T(\theta)}\left|\Expec\left(\check\sigma_{N,T}^2\left(u\right) - \bar\sigma_N^2\right)\right|\to0
	\qquad\text{and}\qquad
		\sup_{u\in B_T(\theta)}\left|\check\sigma_{N,T}^2\left(u\right) - \bar\sigma_N^2\right| \toP 0,
	\]
	where $\mathcal B_T(\theta) = \{v:\lfloor T\theta\rfloor < Tv < \lfloor T\theta\rfloor+1\}$.
\end{theorem}

Theorem~\ref{thm:consistency.check.sigma2} could also be generalised to the case
of weighted least squares regression. This is beyond the scope of the paper
and we consider the more general case only in the simulation study.

\section{Dependent panels}\label{sec:dependent}
In this section, we consider the case where cross-sectional dependence exists
across panels. 
As before, let $N$ denote the number of panels, $T$ the number of observations,
$\mu_i$ the mean of panel $i$ before time $t_0=\lfloor\theta T\rfloor$, $\theta\in(0,1)$,
and $\delta_i$ the size of the change in the mean of panel $i$ at time $t_0+1$.
Following the idea in \citet{bai2002}, we model the cross-sectional dependence using
a common factor model. Specifically, consider
\begin{equation}\label{eq:mainmodel.factors}
		Y_{i,t} = \mu_i + \delta_i \II(t > t_0) + \vec\lambda_i^\top\vec f_t + e_{i,t},
		\quad
		i=1,\ldots,N, \ t=1,\ldots,T,
\end{equation}
where $\vec f_t$ denotes the $p$-vector of common factors at time $t$,
and $\vec\lambda_i=\vec\lambda_{i,N}\in\mathbb{R}^p$ the corresponding
factor loadings associated with panel $i$.

Similar to \citet{horvath2022} and \citet{huskova2024}, we assume that the $\vec\lambda_i$ are bounded
and that the sequence $\{\vec f_t\}$ satisfies a functional central limit theorem.
Formally, the additional assumptions are as follows:
\begin{enumerate}[label=(A\arabic*),start=3,leftmargin=*]
	\item \label{assump.cf.negl}
				$\limsup_{N\to\infty} \max_{1\leq i\leq N}\|\vec\lambda_{i,N}\|<\infty$.
	\item	\label{assumption:commonfactors}
				The common factor sequence $\{\vec f_t; -\infty <t< \infty\}$ is strictly stationary,
				independent of $\{e_{i,t};1 \le i \le N,-\infty <t< \infty\}$, and satisfies
				$\Expec\vec f_t=\vec 0$, $\Expec\vec f_t^{}\vec f_t^\top=\mat I_p$, and
				\[
						\frac{1}{\sqrt T}\sum_{t=1}^{\lfloor Ts\rfloor}\vec f_t
						\toDSkorokhod
						\vec W_{\mat\Sigma}(s),
				\]
				where $\vec W_{\mat\Sigma}(s)\in\mathbb{R}^p$ is a Gaussian process with
				$\Expec\vec W_{\mat\Sigma}(s)=\vec 0$,
				$\Expec\vec W_{\mat\Sigma}(s)\vec W_{\mat\Sigma}(t)=\mat\Sigma\min(s,t)$, and
				$\mat\Sigma$ a positive definite matrix.
\end{enumerate}

Define the quantities
\[
	\mat Q=\lim_{N\to\infty} \left(\sum_{i=1}^N\|\vec\lambda_i\|^2\right)^{-1} \sum_{i=1}^N \vec\lambda_i\vec\lambda_i^\top
	\quad\text{and}\quad
	\bar\lambda_N = \frac{1}{\sqrt N} \sum_{i=1}^N \|\vec\lambda_i\|^2.
\]

\begin{theorem}\label{thm:dependent}
    Suppose Assumptions~\ref{assumption:moments}--\ref{assumption:commonfactors} are satisfied.
		If $\sqrt N T^{-1}\to0$ and $\eta_T=o(TN^{-1/2})$ as $\min(N,T)\to\infty$, then,
    under $H_0$, the following statements hold:
    \begin{enumerate}[label=(\roman*)]
        \item If $\bar\lambda_N\to0$, then
							\[V_{N,T}\left(\cdot;\hat\sigma_{N,T}^2(\vec w_T)\right)\toDSkorokhod \bar\kappa G,\]
							where $G$ is a zero-mean Gaussian process with covariance kernel $\gamma(s,t\,|\,D,h)$,
							with $\gamma$, $D$ and $h$ as defined in \eqref{eq:covkernel} and \eqref{eq:Dh}.
        \item If $\bar\lambda_N\to\infty$, then
							\[
									\frac{1}{{\bar\lambda}_N} V_{N,T}\left(s;\hat\sigma_{N,T}^2(\vec w_T)\right)
									\toDSkorokhod
									\vec B_{\mat\Sigma}^{\top}(s)\mat Q\vec B_{\mat\Sigma}(s)
										+ \bar B\left(\mat Q,\mat\Sigma\right)m(s),
							\]
							where $\vec B_{\mat\Sigma}(s)=\vec W_{\mat\Sigma}(s)-s\vec W_{\mat\Sigma}(1)$ and
							\[
								\bar B\left(\mat Q,\mat\Sigma\right)=\lim_{T\to\infty}\left(\vec m_T^\top\mat W_T^{}\vec m_T^{}\right)^{-1}\sum_{k=1}^{T-1}\vec B_{\mat\Sigma}^{\top}\left(\frac{k}{T}\right)\mat Q\vec B_{\mat\Sigma}\left(\frac{k}{T}\right)m\left(\frac{k}{T}\right)w_{k,T}.
							\]
    \end{enumerate}
\end{theorem}
We consider two specific choices of $\vec w_T$.
If $\vec w_T$ is the $(T-1)$-vector of ones, i.e.\ when the regression in \eqref{eq:regression.estimator}
is ordinary least squares,
\[
	\bar B\left(\mat Q,\mat\Sigma\right)=30\int_0^1\vec B_{\mat\Sigma}^{\top}(s)\mat Q\vec B_{\mat\Sigma}(s)m(s)ds.
\]
For the case considered earlier where $\vec w_T=\II(Tk=\lfloor T\tau\rfloor)$, it follows that
$\bar B(\mat Q,\mat\Sigma)=\vec B_{\mat\Sigma}^{\top}(\tau)\mat Q\vec B_{\mat\Sigma}(\tau)/m(\tau)$.

\section{Simulation study}\label{sec:simulations}

Because the limit null distribution in Theorem~\ref{thm:dependent} depends on nuisance parameters
that need to be estimated, we employ a wild bootstrap algorithm adapted
from \citet{praskova2024} to obtain critical values of the tests.

First, define the centered panels $\hat\eta_{i,t} = Y_{i,t} - \bar Y_{i,T}$
for $i=1,\ldots,N$, and $t=1,\ldots,T$.
Then, by means of the information criteria proposed by \citet{bai2002},
estimate the number of common factors $p$ present in $\hat\eta_{i,t}$,
the common factor sequence $\vec f_t$, and the factor loadings $\vec\lambda_i$.
Denote estimators of these quantities by $\hat p$, $\hat{\vec\lambda}_i$,
and $\hat{\vec f}_t$, respectively, and use these to determine the residuals
$\hat e_{i,t}=\hat\eta_{i,t}-\hat{\bm\lambda}_i^\top\hat{\bm f}_t$.
To generate one bootstrap realization of $\hat V_{N,T}$,
proceed as follows:
\begin{enumerate}
	\item Generate, independently of all other quantities, strictly stationary
				sequences $\{\xi_{i,t},t=1,\dots,T\}$,
				$i=1,\ldots,N$, with $\Expec(\xi_{i,t})=0$ and $\Var(\xi_{i,t})=1$.
	\item For all $(i,t)$, construct the bootstrap errors
				$\tilde e_{i,t}^*=\xi_{i,t}^{}\hat e_{i,t}$.
	\item	Generate a $T\times \hat p$ matrix $\vec f_t^*$ from a multivariate
				normal distribution having the long-run covariance structure of $\hat{\vec f_t}$.
	\item Construct the bootstrap observations
				$Y_{i,t}^* = \hat{\bm\lambda}_i^\top{\bm f}_t^* + \tilde e_{i,t}^*$.
	\item Construct the process $V_{N,T}^*$ according
				to \eqref{eq:VNT} and \eqref{eq:regression.estimator},
				but using the bootstrap observations
				$Y_{i,t}^*$ instead of the original observations $Y_{i,t}$.
\end{enumerate}
Steps 1 to 5 are repeated $B$ times to obtain many bootstrap
realizations $V_{N,T}^{*1},\ldots,V_{N,T}^{*B}$.

In the simulations, we generate each sequence $\{\xi_{i,t},t=1,\dots,T\}$,
$i=1,\ldots,N$, in Step~1 independently from a zero-mean Gaussian process
with $\Cov(\xi_{i,u},\xi_{i,v})=K((u-v)/b_T)$,
where $K(s)=\min(2\max(0,1-|s|),1)$ and $b_T=\log(T)$.

In the model in \eqref{eq:mainmodel3}, we again consider the error sequences
\ref{M1} and \ref{M2} introduced in Section~\ref{numstudy.asymp}.
For the common factors we consider two cases: $\lambda_i=N^{-1/2}$
and $\lambda_i=N^{-1/8}$ for all $i=1,\ldots,N$. The former corresponds
to the case of weak cross-sectional dependence, case (i) in Theorem~\ref{thm:dependent},
and the latter to the case of strong cross-sectional dependence, case (ii).
The common factor sequence $\vec f_t$ is taken to be a univariate ($p=1)$
sequence of i.i.d.\ N$(0,1)$ random variables.

In the tables that follow, we use the notation $\hat V^{\rm{ols}}$,
$\hat V^{\rm{wls}}$, $\hat V^{0.1}$ and $\hat V^{0.5}$ introduced in
Section~\ref{numstudy.asymp} to refer to tests using the mean long-run variance estimator
in \eqref{eq:regression.estimator} with respective weights.
We compare the performance of these tests to the corresponding
tests making use of the alternative estimator introduced in
Section~\ref{sec:improvedestimator},
which are referred to in the tables using the obvious notation
$\check V^{\rm{ols}}$, $\check V^{\rm{wls}}$, $\check V^{0.1}$
and $\check V^{0.5}$. For this numerical study, the weights $\vec w_T$ appearing
in \eqref{eq:regression.estimator} and in the estimator of Section~\ref{sec:improvedestimator}
were chosen to be the same for corresponding tests $\hat V$ and $\check V$.

Table~\ref{tab:H0.weak} shows the rejection percentages in the case of
weak dependence between panels when there is no change in the cross-sectional mean.
Overall, the tests in the left-hand panel seem to be reasonably level-preserving,
except for a few exceedances visible for the test $\hat V^{\rm{wls}}$ which diminish
rapidly as $N$ and $T$ are increased. Recall from Corollary~\ref{thm.wls} that the limit
null distribution of $\hat V^{\rm{wls}}$ relies on the condition $\sqrt NT^{-1}\log T\to0$
as opposed to the weaker condition $\sqrt NT^{-1}\to0$ required in other tests,
which might explain the slower convergence of the empirical level of this test
to the nominal level. Similar observations can be made in the right-hand panel for
tests using the estimator of Section~\ref{sec:improvedestimator} but with more severe
size distortion for smaller sample sizes.

The rejection percentages in the case of changes $\delta_i\sim\rm{U}[-0.4,0.4]$
at time $t_0=\lfloor\frac{1}{2}T\rfloor$
in 50\% of the panels is presented in Table~\ref{tab:HA.weak}. Clearly,
most tests exhibit increasing power with increasing sample size, which
is in agreement with Theorem~\ref{thm:VNT.consistency}. Notice that
the test $\smash{\hat V^{\rm{wls}}}$ has the highest power
among the tests based on $\hat\sigma_{N,T}^2$ in
\eqref{eq:regression.estimator}, with $\smash{\hat V^{0.1}}$ having second-highest power.
However, choosing $\tau=0.5$ as in test $\smash{\hat V^{0.5}}$ has a significant
negative impact on power, highlighting how crucial a proper choice of $\tau$ is.
As expected, the tests in the right-hand panel employing the estimator
$\check\sigma_{N,T}^2$ in
Section~\ref{sec:improvedestimator} all have higher power than their counterparts
based on $\hat\sigma_{N,T}^2$.

We now move on to the case of strong dependence between panels,
the results of which are shown in Tables~\ref{tab:H0.strong}
and \ref{tab:HA.strong}. Again, the empirical size of the tests
shown in Table~\ref{tab:H0.strong} are reasonable,
with a few tests being liberal in some cases.
As can be seen in Table~\ref{tab:HA.strong}, most of the
tests employing the estimator in \eqref{eq:regression.estimator}
seem to have very low power, which is ameliorated
to some extent when using the estimator of Section~\ref{sec:improvedestimator}. 

\begin{table}[htp]
	\centering
	\caption{Rejection percentages
					at the $5\%$ significance level in the case of no change and
					weak cross-sectional dependence.}
	\label{tab:H0.weak}
	\begin{tabular}{lrrrrrrrrrrrrrr}
		\toprule
		&&& \multicolumn{4}{c}{Estimator $\hat\sigma^2$} & \multicolumn{4}{c}{Estimator $\check\sigma^2$} \\
		\cmidrule(lr){4-7}\cmidrule(lr){8-11}
		Model & $N$ & $T$ & $\hat V^{\rm{ols}}$ & $\hat V^{\rm{wls}}$ & $\hat V^{0.1}$ & $\hat V^{0.5}$
											& $\check V^{\rm{ols}}$ & $\check V^{\rm{wls}}$ & $\check V^{0.1}$ & $\check V^{0.5}$ \\
		\midrule
		AR(0) & 50 & 50 & 5.2 & 4.8 & 1.6 & 4.7 &  6.5 &  5.7 & 1.7 &  0.4 \\
		 &  & 100 & 3.4 & 3.9 & 3.4 & 3.9 &  4.2 &  5.2 & 3.6 &  0.6 \\
		 &  & 200 & 3.9 & 4.3 & 5.2 & 6.2 &  5.1 &  6.8 & 3.8 &  0.4 \\
		\cmidrule{2-11}
		 & 100 & 50 & 2.6 & 1.8 & 1.2 & 3.1 &  6.8 &  7.3 & 1.9 &  0.1 \\
		 &  & 100 & 4.4 & 4.3 & 2.2 & 3.4 &  5.0 &  5.9 & 2.6 &  0.1 \\
		 &  & 200 & 4.8 & 5.0 & 3.4 & 4.4 &  6.3 &  6.4 & 4.6 &  0.4 \\
		\cmidrule{2-11}
		 & 200 & 50 & 5.2 & 4.4 & 2.2 & 5.5 &  8.6 & 10.0 & 3.3 &  0.4 \\
		 &  & 100 & 5.0 & 4.4 & 2.5 & 4.0 &  4.9 &  5.8 & 3.6 &  0.4 \\
		 &  & 200 & 7.1 & 6.0 & 4.8 & 4.2 &  6.0 &  6.5 & 5.7 &  0.4 \\
		\midrule
		AR(0.3) & 50 & 50 & 4.7 & 5.3 & 2.6 & 4.0 &  5.5 & 10.0 & 3.7 &  0.9 \\
		 &  & 100 & 3.6 & 4.4 & 4.4 & 3.7 &  4.0 &  4.9 & 4.7 &  0.6 \\
		 &  & 200 & 3.7 & 5.5 & 4.7 & 5.3 &  5.0 &  7.1 & 4.3 &  1.1 \\
		\cmidrule{2-11}
		 & 100 & 50 & 2.2 & 4.9 & 3.5 & 3.1 &  7.5 & 11.9 & 4.6 &  4.3 \\
		 &  & 100 & 4.4 & 4.9 & 2.1 & 3.0 &  4.0 &  7.8 & 3.1 &  0.5 \\
		 &  & 200 & 4.9 & 5.0 & 4.5 & 5.0 &  6.2 &  5.8 & 5.8 &  0.7 \\
		\cmidrule{2-11}
		 & 200 & 50 & 4.1 & 7.6 & 4.6 & 4.1 & 10.6 & 17.8 & 7.5 & 27.9 \\
		 &  & 100 & 3.7 & 4.0 & 3.5 & 2.9 &  5.3 &  7.8 & 4.4 &  8.7 \\
		 &  & 200 & 5.6 & 5.4 & 5.0 & 4.1 &  5.3 &  6.0 & 6.0 &  3.4 \\
		\midrule
		ARMA & 50 & 50 & 6.0 & 5.8 & 0.6 & 6.1 &  7.6 & 10.7 & 0.8 &  3.4 \\
		 &  & 100 & 3.5 & 4.9 & 2.5 & 4.1 &  4.8 &  8.9 & 3.1 &  2.3 \\
		 &  & 200 & 4.5 & 5.3 & 4.4 & 6.3 &  7.0 &  7.2 & 4.1 &  3.4 \\
		\cmidrule{2-11}
		 & 100 & 50 & 4.3 & 5.4 & 0.3 & 4.2 &  8.2 & 14.9 & 0.8 &  3.9 \\
		 &  & 100 & 5.7 & 6.0 & 1.4 & 4.0 &  5.9 &  9.7 & 2.3 &  3.8 \\
		 &  & 200 & 5.5 & 6.3 & 3.8 & 5.1 &  7.1 & 10.0 & 4.2 &  4.3 \\
		\cmidrule{2-11}
		 & 200 & 50 & 5.4 & 8.3 & 0.9 & 6.1 & 11.6 & 24.6 & 2.8 & 10.3 \\
		 &  & 100 & 4.7 & 6.3 & 1.6 & 4.8 &  6.3 & 11.9 & 2.5 & 11.1 \\
		 &  & 200 & 7.6 & 7.5 & 3.3 & 4.2 &  5.8 & 10.7 & 5.5 &  7.2 \\
		\bottomrule
	\end{tabular}
\end{table}

\begin{table}[htp]
	\centering
	\caption{Rejection percentages
					at the $5\%$ significance level
					under weak cross-sectional dependence with changes occurring in 50\% of
					the panels.}
	\label{tab:HA.weak}
	\begin{tabular}{lrrrrrrrrrrrrrr}
		\toprule
		&&& \multicolumn{4}{c}{Estimator $\hat\sigma^2$} & \multicolumn{4}{c}{Estimator $\check\sigma^2$} \\
		\cmidrule(lr){4-7}\cmidrule(lr){8-11}
		Model & $N$ & $T$ & $\hat V^{\rm{ols}}$ & $\hat V^{\rm{wls}}$ & $\hat V^{0.1}$ & $\hat V^{0.5}$
											& $\check V^{\rm{ols}}$ & $\check V^{\rm{wls}}$ & $\check V^{0.1}$ & $\check V^{0.5}$ \\
		\midrule
		AR(0) & 50 & 50 &  2.2 &   7.9 &   8.4 &  0.4 &   9.1 &  13.5 &  10.8 &  0.0 \\
		 &  & 100 &  4.0 &  24.4 &  23.2 &  0.1 &  24.9 &  41.6 &  29.4 &  0.0 \\
		 &  & 200 & 15.5 &  63.9 &  64.0 &  0.0 &  75.6 &  88.0 &  72.3 &  0.0 \\
		\cmidrule{2-11}
		 & 100 & 50 &  4.3 &  14.8 &  11.1 &  0.0 &  22.5 &  32.9 &  18.0 &  0.0 \\
		 &  & 100 & 24.0 &  59.1 &  48.5 &  2.0 &  67.5 &  79.4 &  56.3 &  0.0 \\
		 &  & 200 & 68.3 &  97.1 &  93.8 &  7.1 &  99.7 &  99.8 &  97.9 &  0.0 \\
		\cmidrule{2-11}
		 & 200 & 50 & 17.6 &  37.9 &  28.0 &  9.1 &  55.9 &  67.0 &  42.3 &  0.1 \\
		 &  & 100 & 63.9 &  91.1 &  81.0 & 27.3 &  95.6 &  98.8 &  90.0 &  1.9 \\
		 &  & 200 & 99.5 & 100.0 & 100.0 & 87.9 & 100.0 & 100.0 & 100.0 & 29.0 \\
		\midrule
		AR(0.3) & 50 & 50 &  2.3 &   7.6 &   6.7 &  1.5 &   7.1 &  20.6 &   8.6 &  0.6 \\
		 &  & 100 &  1.7 &  10.7 &  13.3 &  0.2 &   9.0 &  23.6 &  16.0 &  0.0 \\
		 &  & 200 &  4.4 &  30.8 &  27.2 &  0.0 &  30.1 &  53.0 &  35.6 &  0.0 \\
		\cmidrule{2-11}
		 & 100 & 50 &  2.2 &  13.9 &   9.4 &  0.7 &  15.9 &  33.5 &  13.1 & 12.3 \\
		 &  & 100 &  6.5 &  26.1 &  17.3 &  0.5 &  26.7 &  50.0 &  23.1 &  7.6 \\
		 &  & 200 & 22.7 &  61.8 &  56.3 &  3.0 &  70.4 &  82.6 &  67.4 & 15.4 \\
		\cmidrule{2-11}
		 & 200 & 50 &  7.4 &  29.3 &  18.6 &  5.3 &  39.0 &  58.7 &  29.5 & 64.8 \\
		 &  & 100 & 16.4 &  52.8 &  40.1 &  6.4 &  62.3 &  79.7 &  49.3 & 67.2 \\
		 &  & 200 & 66.9 &  95.3 &  87.8 & 39.5 &  97.3 &  99.8 &  93.8 & 87.8 \\
		\midrule
		ARMA & 50 & 50 &  2.5 &   4.3 &   2.5 &  1.2 &   5.1 &   7.1 &   5.5 &  0.5 \\
		 &  & 100 &  2.1 &  15.0 &  15.7 &  0.1 &  13.9 &  21.2 &  19.9 &  0.0 \\
		 &  & 200 & 11.6 &  48.9 &  52.8 &  0.0 &  62.7 &  74.0 &  59.6 &  0.0 \\
		\cmidrule{2-11}
		 & 100 & 50 &  3.1 &   4.7 &   3.9 &  0.5 &   7.7 &  10.5 &   6.0 &  0.3 \\
		 &  & 100 & 17.1 &  38.2 &  30.9 &  1.0 &  43.2 &  49.9 &  40.8 &  0.0 \\
		 &  & 200 & 53.3 &  88.9 &  85.2 &  5.2 &  95.6 &  98.2 &  91.1 &  0.0 \\
		\cmidrule{2-11}
		 & 200 & 50 &  8.9 &  11.5 &  10.2 &  3.6 &  21.6 &  22.4 &  14.5 &  0.8 \\
		 &  & 100 & 41.9 &  71.2 &  57.6 & 17.3 &  83.3 &  85.8 &  72.3 &  0.0 \\
		 &  & 200 & 96.9 &  99.9 &  99.5 & 71.9 & 100.0 & 100.0 &  99.9 &  0.7 \\
		\bottomrule
	\end{tabular}
\end{table}

\begin{table}[htp]
	\centering
	\caption{Rejection percentages
					at the $5\%$ significance level in the case of no change and
					strong cross-sectional dependence.}
	\label{tab:H0.strong}
	\begin{tabular}{lrrrrrrrrrrrrrr}
		\toprule
		&&& \multicolumn{4}{c}{Estimator $\hat\sigma^2$} & \multicolumn{4}{c}{Estimator $\check\sigma^2$} \\
		\cmidrule(lr){4-7}\cmidrule(lr){8-11}
		Model & $N$ & $T$ & $\hat V^{\rm{ols}}$ & $\hat V^{\rm{wls}}$ & $\hat V^{0.1}$ & $\hat V^{0.5}$
											& $\check V^{\rm{ols}}$ & $\check V^{\rm{wls}}$ & $\check V^{0.1}$ & $\check V^{0.5}$ \\
		\midrule
		AR(0) & 50 & 50 &  7.4 &  8.6 &  5.3 &  6.6 &  7.4 &  9.4 &  5.6 &  6.1 \\
		 &  & 100 &  8.0 &  8.9 &  6.5 &  7.5 &  6.2 &  7.7 &  6.7 &  5.5 \\
		 &  & 200 &  9.0 & 11.1 &  7.5 &  8.6 &  8.1 &  8.9 &  7.6 &  6.1 \\
		\cmidrule{2-11}
		 & 100 & 50 &  9.2 &  8.6 &  6.1 &  8.0 &  7.8 &  8.0 &  7.9 &  9.1 \\
		 &  & 100 &  9.3 &  9.9 &  8.7 &  8.2 &  7.5 &  8.3 &  6.7 &  7.4 \\
		 &  & 200 & 13.3 & 13.0 &  9.0 & 11.2 &  8.5 &  8.5 &  9.7 &  8.8 \\
		\cmidrule{2-11}
		 & 200 & 50 &  9.4 & 11.4 &  9.1 &  8.3 &  9.4 & 10.2 &  8.0 &  9.7 \\
		 &  & 100 & 10.8 & 10.2 &  9.6 &  8.3 &  7.7 &  8.6 &  7.6 &  9.6 \\
		 &  & 200 & 14.7 & 15.2 & 11.4 & 12.8 &  8.8 & 10.3 & 10.9 &  9.2 \\
		\midrule
		AR(0.3) & 50 & 50 &  3.1 &  4.2 &  3.5 &  2.2 &  4.6 &  5.6 &  3.4 &  5.8 \\
		 &  & 100 &  4.1 &  4.8 &  4.4 &  4.2 &  3.7 &  4.9 &  2.9 &  4.4 \\
		 &  & 200 &  5.8 &  5.5 &  5.4 &  4.9 &  4.4 &  6.0 &  4.1 &  3.2 \\
		\cmidrule{2-11}
		 & 100 & 50 &  2.8 &  4.3 &  5.6 &  1.7 &  4.7 &  7.8 &  4.3 & 14.1 \\
		 &  & 100 &  4.6 &  5.3 &  4.7 &  3.6 &  5.6 &  7.0 &  4.7 & 10.4 \\
		 &  & 200 &  7.0 &  6.7 &  4.8 &  5.2 &  4.6 &  5.3 &  5.3 &  7.1 \\
		\cmidrule{2-11}
		 & 200 & 50 &  4.4 &  5.1 &  6.3 &  3.7 &  4.7 &  6.7 &  6.5 & 20.8 \\
		 &  & 100 &  4.9 &  6.9 &  6.9 &  4.9 &  4.9 &  6.9 &  7.6 & 21.3 \\
		 &  & 200 &  6.2 &  7.5 &  8.3 &  6.3 &  5.5 &  8.4 &  8.0 & 20.7 \\
		\bottomrule
	\end{tabular}
\end{table}

\begin{table}[htp]
	\centering
	\caption{Rejection percentages
					at the $5\%$ significance level
					under strong cross-sectional dependence with changes occurring in 50\% of
					the panels.}
	\label{tab:HA.strong}
	\begin{tabular}{lrrrrrrrrrrrrrr}
		\toprule
		&&& \multicolumn{4}{c}{Estimator $\hat\sigma^2$} & \multicolumn{4}{c}{Estimator $\check\sigma^2$} \\
		\cmidrule(lr){4-7}\cmidrule(lr){8-11}
		Model & $N$ & $T$ & $\hat V^{\rm{ols}}$ & $\hat V^{\rm{wls}}$ & $\hat V^{0.1}$ & $\hat V^{0.5}$
											& $\check V^{\rm{ols}}$ & $\check V^{\rm{wls}}$ & $\check V^{0.1}$ & $\check V^{0.5}$ \\
		\midrule
		AR(0) & 50 & 50 & 4.0 &  7.3 &  6.6 & 2.2 &  7.7 & 11.4 &  6.9 &  5.5 \\
		 &  & 100 & 1.8 &  6.8 & 10.3 & 0.7 & 10.7 & 12.8 & 11.9 &  8.0 \\
		 &  & 200 & 0.6 &  7.4 & 16.2 & 0.1 & 20.2 & 25.6 & 25.2 & 20.1 \\
		\cmidrule{2-11}
		 & 100 & 50 & 4.0 &  7.1 &  8.6 & 2.9 &  9.2 & 10.0 & 10.2 & 10.1 \\
		 &  & 100 & 1.4 &  7.5 & 13.3 & 0.6 & 11.6 & 13.0 & 13.6 & 10.6 \\
		 &  & 200 & 1.1 &  8.8 & 24.9 & 0.1 & 29.6 & 35.7 & 39.6 & 38.8 \\
		\cmidrule{2-11}
		 & 200 & 50 & 5.3 & 10.0 & 11.8 & 2.1 & 11.6 & 13.6 & 11.6 &  9.2 \\
		 &  & 100 & 2.8 &  9.2 & 17.2 & 0.7 & 13.8 & 16.5 & 15.9 & 15.6 \\
		 &  & 200 & 1.3 & 13.7 & 38.7 & 0.0 & 50.6 & 62.5 & 58.9 & 62.6 \\
		\midrule
		AR(0.3) & 50 & 50 & 1.5 &  3.8 &  4.4 & 1.2 &  4.9 &  7.0 &  5.1 &  8.5 \\
		 &  & 100 & 2.0 &  4.8 &  5.7 & 1.4 &  6.4 &  9.5 &  4.7 & 10.2 \\
		 &  & 200 & 1.5 &  6.0 & 12.8 & 0.1 & 15.8 & 23.0 & 14.3 & 23.4 \\
		\cmidrule{2-11}
		 & 100 & 50 & 2.2 &  4.3 &  6.8 & 0.6 &  6.3 & 10.5 &  6.2 & 23.4 \\
		 &  & 100 & 2.1 &  5.5 &  7.3 & 0.3 &  9.3 & 12.1 &  8.9 & 31.8 \\
		 &  & 200 & 0.9 &  9.4 & 13.6 & 0.1 & 18.6 & 25.2 & 21.7 & 53.9 \\
		\cmidrule{2-11}
		 & 200 & 50 & 2.6 &  5.0 &  8.0 & 2.2 &  6.3 &  8.7 &  9.7 & 39.7 \\
		 &  & 100 & 1.8 &  8.1 & 14.3 & 0.4 &  8.9 & 13.9 & 16.1 & 68.1 \\
		 &  & 200 & 2.0 & 12.9 & 32.1 & 0.2 & 36.4 & 53.3 & 39.5 & 94.0 \\
		\bottomrule
	\end{tabular}
\end{table}

\section{Conclusion}\label{sec:conclusions}

In this paper, we proposed a class of change-point tests
for high-dimensional panel data exhibiting temporal and
cross-sectional dependence. The asymptotic null distribution of the test
process was derived and it was shown that the test is consistent
under the alternative.
Most of the test were demonstrated to have favorable finite-sample properties,
both in terms of empirical size and power. Generally, tests based
on the newly proposed estimators of the mean long-run variance of
panels outperform existing tests. In addition, it was shown that adjusting
the long-run variance estimator for a potential change-point
improves overall finite-sample performance.

\begin{acknowledgement}
The first author expresses his deep gratitude to Professor Marie Hu\v{s}kov\'{a}
for her kind-hearted mentorship and insightful collaboration.
Her guidance has been instrumental in shaping his academic development
and will remain a lasting influence on his future work.
The author also thanks the editor and the anonymous reviewer for their
constructive comments and suggestions, which certainly helped to improve
the quality of the manuscript.
The research of the second author was partially funded by the
National Research Foundation of South Africa (Grant Number 127908).
\end{acknowledgement}

\ethics{Competing Interests}{%
The authors have no conflicts of interest to declare that are relevant to the content of this chapter.}

\eject

\section*{Appendix}\label{appendixA}
\addcontentsline{toc}{section}{Appendix}

\begin{proof}[Proof of Theorem~\ref{thm1}]
	For ease of notation, define
	$\hat V_{N,T}(s)=V_{N,T}(s;\hat\sigma_{N,T}^2(\vec w_T))$
	and
	$\beta_2=\vec m_T^\top\mat W_T^{}\vec m_T^{}$.
	Also define $\bar{\vec z}_{N,T}$ as the $(T-1)$-vector with $k$th entry $N^{-1}\sum_{i=1}^N\bar Z_{i,T}^2(k/T)$,
	where
	\[\bar Z_{i,T}(s)
		= \frac{1}{\sqrt T}\sum_{k=1}^{\lfloor Ts \rfloor}\left( e_{i,k} - \bar e_{i,T} \right).\]
	Under $H_0$ and the conditions of Theorem~\ref{thm1},
	\[\begin{split}
		\hat\sigma_{N,T}^2(\vec w_T) - \bar\sigma_N^2
		&= \beta_2^{-1}\vec m_T^\top\mat W_T^{}\left(\bar{\vec z}_{N,T} - \Expec\bar{\vec z}_{N,T}\right) - \beta_2^{-1}\vec m_T^\top\mat W_T^{}\left(\Expec{\bar{\vec z}}_{N,T} - \bar\sigma_N^2\vec m_T\right) \\
		&= \beta_2^{-1}\vec m_T^\top\mat W_T^{}\dot{\vec z}_{N,T} + \eta_TO\left(\frac{1}{T}\right),
	\end{split}\]
	where $\dot{\vec z}_{N,T}=\bar{\vec z}_{N,T}-\Expec\bar{\vec z}_{N,T}$.
	Here we made use of the fact that, under the imposed weak dependence conditions,
	$\max_{1\le i\le N}\sup_{s\in(0,1)}|\Expec\bar Z_{i,T}^2(s)-\sigma_i^2m_T(s)|=O(T^{-1})$;
	see, e.g., \citet{huskova2024}.
	Therefore, the process $\hat{V}_{N,T}$ has the representation
	\[\begin{split}
		\hat{V}_{N,T}(s)
		={}&{} \frac{1}{\sqrt N}\sum_{i=1}^N \left(U_{i,T}(s) - \beta_2^{-1}\vec m_T^\top\mat W_T^{}\dot{\vec z}_{N,T} m_T(s)\right)
				+ \eta_TO\left(\frac{\sqrt N}{T}\right) + O\left(\frac{\sqrt N}{T}\right) \\
		={}&{} \frac{1}{\sqrt N}\sum_{i=1}^N \left(U_{i,T}(s) - \beta_2^{-1}\vec m_T^\top\mat W_T^{}\dot{\vec z}_{N,T} m_T(s)\right)
				+ o(1),
	\end{split}\]
	uniformly in $s\in(0,1)$ as $\min(N,T)\to\infty$, where
	$U_{i,T}(s)=\bar Z^2_{i,T}(s) - \Expec \bar Z^2_{i,T}(s)$.
	Using the Lyapounov condition for convergence together with the Cram\'er--Wold device
	as in \citet{horvath2012},
	it can be shown that the finite-dimensional distributions of $\hat{V}_{N,T}$ are
	asymptotically normal if $\nu>4$.
	
	It can be shown that $\max_{1\le i\le N}\sup_{0<s\le t<1}|\Cov(U_{i,T}(s),U_{i,T}(t))-2\sigma_i^4g_T^2(s,t)|=O(T^{-1})$;
	see, for example, \citet{huskova2024}. Consequently, $\Expec[\dot{\vec z}_{N,T}^{}\dot{\vec z}_{N,T}^\top]=2N^{-1}\bar\kappa_N^2\mat C_T + O(N^{-1}T^{-1})$,
	so that, due to independence of the panels,
	\[\begin{split}
		\Var\left(\beta_2^{-1}\vec m_T^\top\mat W_T^{}\dot{\vec z}_{N,T}\right)
		&= \frac{1}{\beta_2^2}\vec m_T^\top\mat W_T^{}\Expec\left[\dot{\vec z}_{N,T}^{}\dot{\vec z}_{N,T}^\top\right]\mat W_T^{}\vec m_T \\
		&= 2\bar\kappa_N^2\frac{1}{N\beta_2^2}\vec m_T^\top\mat W_T^{}\mat C_T\mat W_T^{}\vec m_T + O\left(\frac{\eta_T}{NT}\right)
		= \frac{2}{N}\bar\kappa_N^2D_T + o\left(1\right).
	\end{split}\]
	Similarly,
	\[\begin{split}
		\Cov\left(U_{i,T}(s),\beta_2^{-1}\vec m_T^\top\mat W_T^{}\dot{\vec z}_{N,T}\right)
		&= 2\bar\kappa_N^2\frac{1}{\beta_2}\sum_{k=1}^{T-1}g_T\left(s,\frac{k}{T}\right)m_T\left(\frac{k}{T}\right)w_{k,T}^{}
			+ O(T^{-1}) \\
		&=: 2\bar\kappa_N^2h_T(s;\bm w_T) + o(1),
	\end{split}\]
	uniformly in $s\in(0,1)$. Combining these results, we obtain, uniformly in $s,t\in(0,1)$,
	\[\begin{split}
		&\quad\Cov\left(\hat{V}_{N,T}(s),\hat{V}_{N,T}(t)\right) \\
		&= 2\bar\kappa_N^2\left[g_T^2(s,t) + D_T(\vec w_T)m(s)m(t) + h_T(s;\vec w_T)m(t) + h_T(t;\vec w_T)m(s)\right]
				+ o(1) \\
		&\to \bar\kappa^2\gamma(s,t\,|\,D,h).
	\end{split}\]
	
	We now show that the process $\hat{V}_{N,T}$ is tight in $\mathcal D[0,1]$.
	Fix $s,t\in(0,1)$. In what follows, $d_\ell>0$ denotes a general constant
	neither depending on $N$ nor $T$.
	Following the same steps as in \citet{horvath2012}, one can show that
	$\Expec|N^{-1/2}\sum_{i=1}^N(U_{i,T}(s)-U_{i,T}(t))|^{\nu/2}\le d_1|s-t|^{\nu/4}$.
	Write
	\[
		\beta_2^{-1}\vec m_T^\top\mat W_T^{}\dot{\vec z}_{N,T}
		=\beta_2^{-1}\frac{1}{N}\sum_{i=1}^N\sum_{k=1}^{T-1}m_kw_k\left[\bar Z_{i,T}^2\left(\frac{k}{T}\right)-\Expec\bar Z_{i,T}^2\left(\frac{k}{T}\right)\right]
		=:\frac{1}{N}\sum_{i=1}^NW_{i,T}.
	\]
	Fix $\lambda\in[2,\nu/2)$.
	Observe that, since the $W_{i,T}$ are zero-mean independent random variables and
	\[
		\Expec|W_{i,T}|^\gamma
		\le d_2\beta_2^{-\gamma}(T-1)^{\gamma-1}\sum_{k=1}^{T-1}(m_kw_k)^\gamma\Expec\left|\bar Z_{i,T}^2\left(\frac{k}{T}\right)-\Expec\bar Z_{i,T}^2\left(\frac{k}{T}\right)\right|^\gamma
		= O(1),
	\]
	it follows by the Rosenthal inequality that
	\[\begin{split}
		\Expec\left|\sum_{i=1}^N W_{i,T}\right|^{\gamma}
		\le d_3\left\{\sum_{i=1}^N \Expec\left|W_{i,T}\right|^{\gamma} 
									+ \left(\sum_{i=1}^N \Expec\left|W_{i,T}\right|^2\right)^{\gamma/2} \right\}
		\le d_4 N^{\gamma/2}.
	\end{split}\]
	Therefore, since $|m_T(s) - m_T(t)|^{\gamma} \le d_5|s-t|^{\gamma} + d_6T^{-\gamma}$, we have
	\[\begin{split}
		\Expec\left|\frac{1}{\sqrt N}\sum_{i=1}^N W_{i,T}\left(m_T(s) - m_T(t)\right)\right|^{\gamma}
		= \Expec\left|\frac{1}{\sqrt N}\sum_{i=1}^N W_{i,T}\right|^{\gamma}\left|m_T(s) - m_T(t)\right|^{\gamma}
		\le d_7|s-t|^{\gamma},
	\end{split}\]
	whence it follows that
	$\Expec|\hat{V}_{N,T}(s) - \hat{V}_{N,T}(t)|^{\nu/2}\le d_8|s-t|^{\nu/2}$
	and, since $\nu>4$, tightness follows by Theorem~12.3 of \citet[p.~95]{billingsley1968}.
\end{proof}

\begin{proof}[Proof of Theorem~\ref{thm:VNT.consistency}]
Suppose that there is a change point at $\theta$ and write
\[
    V_{N,T}(\theta;\hat\sigma_{N,T}^2)
    = V_{N,T}(\theta;\bar\sigma_N^2) + \sqrt{N}(\hat\sigma_{N,T}^2-\bar\sigma_N^2)m_T(\theta).
\]
Under the stated assumptions, it can be shown \citep[cf.][]{horvath2012} that
\[
    V_{N,T}(\theta;\bar\sigma_N^2) = m_T^2(\theta)\frac{T}{\sqrt{N}}\sum_{i=1}^N\delta_i^2
        + o_{\Prob}\left(\frac{T}{\sqrt{N}}\sum_{i=1}^N\delta_i^2\right).
\]
One can also show that
\begin{equation}\begin{split}\label{consistency.term2}
\sqrt{N}(\hat\sigma_{N,T}^2-\bar\sigma_N^2)
={}&{} \sqrt N\beta_2^{-1}\vec m_T^\top\mat W_T^{}\bar{\vec z}_{N,T} + h(\theta;\vec w_T)\left(\frac{T}{\sqrt{N}}\sum_{i=1}^{N}\delta_i^2\right)
		+ o_{\Prob}\left(\frac{T}{\sqrt{N}}\sum_{i=1}^N\delta_i^2\right) \\
={}&{} h_T(\theta;\vec w_T)\left(\frac{T}{\sqrt{N}}\sum_{i=1}^{N}\delta_i^2\right)
		+ \eta_TO_{\Prob}\left(\frac{\sqrt N}{T}\right)
		+ o_{\Prob}\left(\frac{T}{\sqrt{N}}\sum_{i=1}^N\delta_i^2\right). \\
\end{split}\end{equation}
Since $h_T(\theta;\vec w_T)$ is positive,
$V_{N,T}(\theta;\bar\sigma_N^2)\toP\infty$.
\end{proof}

The proofs of Theorems~\ref{theorem:consistency_kappa1} and
\ref{thm:consistency.check.sigma2} follow from lengthy but elementary calculations
and are therefore omitted.

\clearpage
\begin{proof}[Proof of Theorem~\ref{thm:dependent}]
	By Lemma~1.2 in the supplement to \citet{horvath2022},
	\[\begin{split}
		&V_{N,T}\left(s;\bar\sigma_N^2\right) \\
		&= \frac{1}{\sqrt N}\sum_{i=1}^N\left(\bar Z_{i,T}^2(s) - \bar\sigma_N^2m_T(s)\right)
				+ \frac{1}{T\sqrt N}\sum_{i=1}^N\left(\vec\lambda_i^\top\sum_{t=1}^{\lfloor Ts\rfloor}\left(\vec f_t-\bar{\vec f}_T\right)\right)^2
				+ o_{\Prob}\left(\bar\lambda_N\right) \\
		&= \frac{1}{\sqrt N}\sum_{i=1}^N\left(\bar Z_{i,T}^2(s) - \bar\sigma_N^2m_T(s)\right)
				+ \frac{1}{\sqrt N}\sum_{i=1}^N\|\vec\lambda_i\|^2\vec B_{\Sigma}^\top(s)\mat Q\vec B_{\Sigma}(s)
				+ o_{\Prob}\left(\bar\lambda_N\right).
	\end{split}\]
	Now consider
	\[\begin{split}
		\sqrt{N}\left(\hat\sigma_{N,T}^2 - \bar\sigma_N^2\right)
		{=}&{} \frac{\sqrt{N}}{\beta_2}\vec m_T^\top\mat W_T^{}\bar{\vec z}_{N,T}
			+ \frac{1}{\beta_2T\sqrt N}\sum_{i=1}^N\sum_{k=1}^{T-1}\left(\vec\lambda_i^\top\sum_{t=1}^k\left(\vec f_t-\bar{\vec f}_T\right)\right)^2m\left(\frac{k}{T}\right)w_{k,T} \\
			{}&{}+ \frac{2}{\beta_2\sqrt{TN}}\sum_{i=1}^N\sum_{k=1}^{T-1}\bar Z_{i,T}\left(\frac{k}{T}\right)\vec\lambda_i^\top\sum_{t=1}^k\left(\vec f_t-\bar{\vec f}_T\right)m\left(\frac{k}{T}\right)w_{k,T} \\
		{=}&{:} \frac{\sqrt{N}}{\beta_2}\vec m_T^\top\mat W_T^{}\bar{\vec z}_{N,T} + A_{N,T} + B_{N,T}.
	\end{split}\]
	Again using Lemma~1.2 of \citet{horvath2022},
	\[\begin{split}
		A_{N,T}
		&= \frac{1}{\sqrt N}\sum_{i=1}^N\|\vec\lambda_i\|^2\frac{1}{\beta_2}\sum_{k=1}^{T-1}\vec B_{\mat\Sigma}^\top\left(\frac{k}{T}\right)\mat Q\vec B_{\mat\Sigma}\left(\frac{k}{T}\right)m\left(\frac{k}{T}\right)w_{k,T}
			+ o_{\Prob}\left(\bar\lambda_N\right),
	\end{split}\]
	and by their Lemma~1.3,
	\[\begin{split}
		B_{N,T}
		&= \frac{2}{\beta_2\sqrt N}\sum_{k=1}^{T-1}m\left(\frac{k}{T}\right)w_{k,T}O_{\Prob}\left(1+\left(\sum_{i=1}^N\|\vec\lambda_i\|^2\right)^{1/2}\right) \\
		&= O_{\Prob}\left(\frac{1}{\sqrt N}\right) + o_{\Prob}\left(\bar\lambda_N^{1/2}\right).
	\end{split}\]
	Combining the expressions above yields
	\[\begin{split}
		V_{N,T}\left(s;\hat\sigma_{N,T}^2\right)
		&= \frac{1}{\sqrt N}\sum_{i=1}^N\left(\bar Z_{i,T}^2(s) - \bar\sigma_N^2m_T(s)\right)
				+ \frac{\sqrt{N}}{\beta_2}\vec m_T^\top\mat W_T^{}\bar{\vec z}_{N,T} \\
				&\quad + \bar\lambda_N\vec B_{\Sigma}^\top(s)\mat Q\vec B_{\Sigma}(s)
				+ \bar\lambda_N\bar B(\mat Q,\mat\Sigma)
				+ o_{\Prob}\left(\bar\lambda_N\right),
	\end{split}\]
	which completes the proof of the theorem.
\end{proof}



\end{document}